\documentclass[aps]{revtex4}
\begin{document} 
\title{Numerical validation of the Kerr metric in Bondi-Sachs form} 
\author{Liebrecht R. Venter and Nigel T. Bishop}
\affiliation{Department of Mathematical Sciences, University of South Africa,
P.O. Box 392, Unisa 0003, South Africa}

\begin{abstract} 
A metric representing the Kerr geometry has been obtained by Pretorius
and Israel. We make a coordinate transformation on this metric,
thereby bringing it into Bondi-Sachs form. In order to validate
the metric, we
evaluate it numerically on a regular grid of the new coordinates.
The Ricci tensor is then
computed, for different discretizations, and found to be
convergent to zero. We also investigate the
behaviour of the metric near the axis of symmetry and confirm
regularity. Finally we investigate a Bondi-Sachs representation
of the Kerr geometry reported by Fletcher and Lun; we confirm
numerically that their metric is Ricci flat, but find that it has
an irregular behaviour at the pole.
\end{abstract}


\maketitle 

\section{Introduction}

It is widely believed that the end-state of most massive objects is a
black hole described by the Kerr geometry. The most powerful source of
gravitational radiation is the merger of a black hole with another
black hole (or other compact object such as a neutron star). Such
processes need to be calculated numerically, and the field of numerical
relativity has been developed to tackle such problems -- see, for
example the review~\cite{lehner}. The most popular approach
to numerical relativity is based on the ADM formalism, and
variations thereof, as discussed in the review cited. In these
formalisms, it is understood how to set initial data for
a Kerr black hole~\cite{cook}.

An alternative approach to numerical relativity is based on an
evolution of the metric variables in the Bondi-Sachs
metric~\cite{bondi,sachs}. The approach has been used by a number of
different groups -- for example~\cite{newnews,particle,roberto,rdi2,siebel1,
bartnik}, and for a review see~\cite{winLR}.
Applications involving black holes have been restricted to the
Schwarzschild case, because of the difficulty in setting initial
and boundary data for a rotating black hole. Also, the
problem of obtaining just the horizon data is much simpler~\cite{win},
and from such data one could then use a numerical code to construct a
complete Bondi-Sachs representation of the Kerr geometry.

The representation of the Kerr geometry obtained by Pretorius and
Israel~\cite{PI-1} (and hereafter this reference will be denoted as PI)
is not in Bondi-Sachs form. However, the PI coordinates are based on
outgoing light cones, and as such should provide a useful
starting-point for the construction of the Bondi-Sachs form of the Kerr
metric. We have constructed a coordinate transformation on the PI
coordinates that brings the transformed metric into Bondi-Sachs form.
In order to validate our metric (and also that of PI) we need to
show that the Ricci tensor vanishes. The metric is not explicit
with respect to the new coordinates, so an analytic approach would
seem to be hopeless. Instead we wrote code that numerically computes the
Ricci tensor for the metric, and confirmed that
the metric is Ricci flat.

In order to describe the Kerr geometry, a metric must not only be Ricci
flat but must also be regular on the axis of symmetry. Thus, we also
find the leading order terms of the metric near the axis, and thereby
confirm regularity.

Fletcher and Lun~\cite{lun} have proposed a metric in Bondi-Sachs form
that represents the Kerr geometry. We will call this the FL metric.
The metric has the convenient
property that its coefficients are explicit functions of its
coordinates, so a numerical evaluation of the Ricci tensor is
straightforward. This we did, confirming that the Ricci tensor vanishes.
However, we also found that it does not have a normal behaviour near
the pole, thus rendering it inappropriate for use in numerical
relativity.
 
Sec.~\ref{s-b} summarizes results obtained in PI that
will be used later. Then in Sec.~\ref{s-prop} we obtain in explicit
form the equations that relate the
Boyer-Lindquist and PI coordinates, and also find explicit expressions
for the PI metric quantities.  
Sec.~\ref{Coordinate transforms to Bondi-Sachs form} derives the
coordinate transformation needed to change the PI metric into
Bondi-Sachs form.
Sec.~\ref{s-n}  discusses the numerical
implementation of the computation of the Ricci tensor, and presents
the results. The properties of the FL metric are discussed in
Sec.~\ref{FletcherLun-metric}. We end with a Conclusion, Sec.~\ref{s-c}.

\section{Background: the PI metric}
\label{s-b}

The Kerr metric in standard Boyer-Lindquist coordinates $(t,r,\theta,\phi)$ is
\begin{equation}
ds^2= \frac{\Sigma}{\Delta}dr^2+\Sigma d\theta^2 +R^2 \sin^2\theta d\phi^2
-\frac{4mar\sin^2\theta}{\Sigma}d\phi dt -\left( 1-\frac{2mr}{\Sigma}\right)dt^2
\end{equation}
where
\begin{equation}
\Sigma=r^2+a^2\cos^2\theta,\; \Delta= r^2+a^2-2mr,\; 
R^2=r^2+a^2+\frac{2ma^2r\sin^2\theta}{\Sigma}.
\end{equation}
The metric obtained by PI
(Eq.~(PI-31)) uses coordinates $(t,r_*,\lambda,\phi)$,
with $r_*=r_*(r,\theta), \lambda=\lambda(r,\theta)$. The metric is 
\begin{equation} 
\label{PI-metric} 
ds^2 = \frac{ \Delta }{R^2}(dr_{*}^2-dt^2)+\frac{L^2}{R^2}d\lambda^2    
   + R^2 \sin^2\theta (d\phi-\omega_B dt)^2,
\end{equation}
where (see Eq.~(PI-30))
\begin{equation}
\omega_B=\frac{2mar}{\Sigma R^2},
\end{equation}
and $L$ is defined in Eq.~(\ref{e-L}) below.
PI often use $\theta_*$ instead of $\lambda$ as a
coordinate, the two quantities being related by
\begin{equation} 
\lambda=\sin^2(\theta_*) 
\end{equation} 
(see Eqs.~(PI-36) and~(PI-37)). 
Thus an alternative form of the metric~(\ref{PI-metric}) is
\begin{equation} 
\label{Bondi-Sachs form of the Kerr-metric-6} 
ds^2 = \frac{ \Delta }{R^2}(dr_{*}^2-dt^2)+\frac{L^2 \sin^2 2\theta_*}{R^2} 
 d\theta_*^2 + R^2 \sin^2\theta (d\phi-\omega_B dt)^2.
\end{equation}
PI define the quantities (see Eqs.~(PI-7))
\begin{equation}
P^2(\theta,\lambda)=a^2(\lambda-\sin^2\theta),\;
Q^2(r,\lambda)=(r^2+a^2)^2-a^2 \lambda \Delta,\;
\end{equation}
and (see Eqs.~(PI-36) and (PI-14))
\begin{equation}
F(r,\theta,\lambda)=\int^\infty_r \frac{dr^\prime}{Q(r^\prime,\lambda)}
   -\int^{\theta_*}_\theta \frac{d\theta^\prime}{P(\theta^\prime,\lambda)}
\mbox{ subject to the constraint } F=0,
\label{e-F}
\end{equation}
as well as (see Eqs.~(PI-28) and (PI-25))
\begin{equation}
L= \mu P Q \mbox{ with } \mu=-\frac{\partial F}{\partial \lambda}.
\label{e-L}
\end{equation}
Eq.~(PI-15), and using Eqs.~(PI-9) and (PI-39), defines $r_*$
\begin{equation}
r_*= \int \frac{r^2+a^2}{\Delta(r)}dr 
  + \int^\infty_r \frac{r^{\prime 2} 
      +a^2-Q(r^\prime,\lambda)}{\Delta(r^\prime)} dr^\prime 
  +\int^\theta_{\theta_*} P(\theta^\prime,\lambda) d \theta^\prime
\label{e-r*}
\end{equation}

There is one minor difference in the notation used in PI and here. 
We regard the mass of the black hole $m$ and the normalized 
angular momentum $a$ as parameters: thus we do not show $a$ or $m$
as variables in any functional dependence list.

In order to evaluate the metric, the implicit equations need to be
reduced to a numerically solvable form. This is now done. 

\section{Properties of the coordinate transformation and metric variables}
\label{s-prop}
\subsection{Relation between $(r,\theta)$ and $(r_*, \theta_*)$}

Eq.~(\ref{e-F}) implies 
\begin{equation}
\label{Quasi_Spherical_Surface_eq} 
\int_r^{\infty}{\frac{dr'}{Q(r',\lambda)}} - 
\int_\theta^{\theta_*}{\frac{d\theta'}{P(\theta',\lambda)}} =0, 
\end{equation} 
which constitutes a relationship of the form function$(r,\theta,\theta_*)=0$. 
We found an  exact solution for this equation but it was more efficient to 
solve the $r$-integral in Eq.(\ref{Quasi_Spherical_Surface_eq}) numerically and 
the $\theta$-integral exactly as, 
\begin{eqnarray} 
\int_\theta^{\theta_*}{\frac{d\theta'}{P(\theta',\lambda)}}&=& 
  \frac{1}{a}\left[\Phi(\theta_*,\theta_*)- \Phi(\theta,\theta_*)\right]\\ 
\Phi(\alpha,\theta_*)&=&\epsilon(\cos\alpha)\Xi\left(\frac{\sin\alpha} 
  {\sin\theta_*},\sin\theta_*\right)\\ 
\Xi(z,k)&=&\int_0^z \frac{dt}{\sqrt{1-t^2} \sqrt{1-k^2t^2}}
\label{IP_}
\end{eqnarray} 
with the condition that $a\neq0$ and where $\epsilon(x)$ is the $signum$ of $x$. 
 
Eq.~(\ref{e-r*}) can be written as 
\begin{equation}
r_* = I_1(r) + I_2(r,\lambda) + I_3(\theta,\lambda)
  \label{eq-PI-15} 
  \end{equation} 
and we now evaluate the integrals $I_1$, $I_2$ and $I_3$. 
\begin{equation} 
I_1=r+ m \ln(r^2+a^2-2mr) + \frac{2 m^2}{\sqrt{\nu}} 
                            Q_0\left( \frac{\sqrt{\nu}}{m-r} \right) + C, 
\end{equation} 
where $\nu=m^2-a^2$; $Q_0(x)$ is the zero-order special Legendre function of
the second kind; $C$ is an arbitrary constant of integration, and we set $C=0$.
$I_2$ is expressed as 
\begin{equation} 
\label{I2} 
I_2=\int_r^\infty{\frac{\zeta^2+a^2-\sqrt{(\zeta^2+a^2)^2-a^2\lambda 
   \Delta(\zeta)} }{\Delta(\zeta)}d\zeta} 
\end{equation} 
and will be evaluated numerically. $I_3$ reduces to 
\begin{equation} 
  I_3=a \left[  
  [E\left(\gamma,t  \right)-E\left(\frac{\pi}{2},t  \right) ]  
  - (1-t^2) \left[ F_E\left(\gamma,t  \right)-F_E\left(\frac{\pi}{2},t \right) \right] 
  \right] 
\end{equation} 
\begin{equation}   
\gamma=\sin^{-1}\frac{\sin\theta}{\sin\theta_*}, \;\;\;t=\sin\theta_* 
\end{equation} 
and where $E$ and $F_E$ are Legendre elliptic integrals of the first and second 
kinds (we use the notation $F_E$, rather than the usual $F$, to avoid any
confusion with the $F$ introduced by PI and defined in Eq.~(\ref{e-F})).

\subsection{The metric variable $L$} 
The function $L$ in the metric needs to be written in a form that can be 
evaluated explicitly. Now, $L=\mu PQ$ and the problem is to find an explict 
expression for $\mu$. Using Eqs.~(\ref{e-F}) and (\ref{e-L})
\begin{equation} 
\mu=T_1+T_2= 
\int_r^\infty{\frac{\frac{\partial}{\partial\lambda}Q(r^\prime,\lambda)} 
{Q(r^\prime,\lambda)^2}dr^\prime} 
+\frac{\partial}{\partial\lambda}\int_\theta^{\sin^{-1} 
 \sqrt{\lambda}}\frac{d\theta}{P(\theta,\lambda)}. 
\end{equation} 
The first term $T_1$ reduces to 
\begin{equation}\label{mu_integral term} 
T_1=-\frac{1}{2}\int_r^\infty{\frac{a^2(r^{\prime 2}+a^2-2mr^\prime)} 
  {\left[(r^{\prime 2}+a^2)^2-a^2\lambda(r^{\prime 2}+a^2-2mr^\prime) 
  \right]^{3/2}}}dr^\prime 
\end{equation} 
and this term will need to be evaluated numerically. 
The second term $T_2$ can be simplified to  
\begin{equation} 
T_2=\frac{1}{2}\frac{    a\sin\theta \cos\theta  }               
     {  a^2\lambda(\lambda-1)\sqrt{\lambda-\sin^2\theta}   
      } 
+\frac{1}{2}\frac{\Omega(\theta)\epsilon(\cos\theta)-\Omega(\theta_*)}  {  
 a^2\lambda(\lambda-1)\sqrt{\lambda-\sin^2\theta} } 
 \end{equation} 
where 
\begin{equation} 
\Omega(\alpha) = a 
\left[E\left(\eta,t  \right)- (1-t^2) F_E\left(\eta,t  \right)\right], \;
\eta= \sin^{-1}\frac{\sin\alpha}{\sin\theta_*},\;
t={\sin\theta_*}.
\end{equation} 

\subsection{Regularity on the axis $\theta_*=0$}
\label{BV_metric4smallTheta}
The Kerr geometry is axially symmetric, and therefore the metric
needs to satisfy a regularity condition on the axis $\theta_*=0$.
In order to check the condition we need to know the values on the
axis of the non-zero metric components, as well as the leading
term in a series expansion in $\theta_*$ for those components that
are zero on the axis. Clearly the following metric functions have
non-zero limits at $\theta_*=\theta=0$: $\Delta, \Sigma, R^2, \omega_B$;
but we will need to work out $Q$ and $P$ near the axis so that we can
obtain expressions for $\theta$ and $L$ near the axis.

We assume that $\theta_*$ and $\theta$ are small so that
\begin{equation}
P =  a \sqrt { \theta_*^{2}-\theta^{2} } 
\end{equation}
and
\begin{equation}
Q  =   \left( {r}^{2}+{a}^{2} \right) 
\left( 1-{\frac {{a}^{2}\theta_*^{2}{\Delta}}{2 {r}^{2}+2 {a}^{2}}}
\right),\;\;
Q^{-1}  =   \left( {r}^{2}+{a}^{2} \right)^{-1}
\left( 1+{\frac {{a}^{2}\theta_*^{2}{\Delta}}{2 {r}^{2}+2 {a}^{2}}}
\right).
\end{equation}
Then Eq. (\ref{Quasi_Spherical_Surface_eq}) leads to
\begin{equation}\label{thetafromTheta}
\theta=  \frac {r  \theta_*}{\sqrt {\left( r^2+a^2 \right)}}.
\end{equation}
Next, we evaluate $\mu$ as
\begin{equation}
\mu=-\frac{\partial F}{\partial \lambda}
=-\frac{1}{2\theta_*}\frac{\partial F}{\partial \theta_*}
=\frac{r}{2a^2\theta_*^2},
\end{equation}
and then $L=\mu P Q $ is 
\begin{equation}
L= \frac{r\sqrt{r^2+a^2}}{2\theta_*}.
\end{equation}

Thus the non-trivial metric components at, or near, the $\theta_*=0$ axis can
be evaluated. We find that the PI metric satisfies
\begin{equation}
g_{\theta_*\theta_*}=r^2,\;\; g_{\phi\phi}=r^2\theta_*^2,\;\;
g_{t\phi}=-\frac{2 r^3 \theta_*^2 m a}{(r^2+a^2)^2}.
\end{equation}
Since $\theta_*^2 g_{\theta_*\theta_*}/g_{\phi\phi}\rightarrow 1$
as $\theta_*\rightarrow 0$, the metric is regular on the axis.

\section{Coordinate transformation to Bondi-Sachs form}
\label{Coordinate transforms to Bondi-Sachs form}

The transformation
\begin{equation}\label{coord_xform1}
t  \rightarrow u=t-r_*
\end{equation}
brings the metric (\ref{PI-metric}) into Bondi-Sachs form
except for the last term
\begin{equation}
R^2 sin^2\theta(d\phi-\omega_B(du+dr_*))^2;
\end{equation}
but this term can be brought into the required form with the transformation
\begin{equation}\label{coord_xform2}
\phi \rightarrow \phi_* = \phi+H(r_*,\theta_*)
\end{equation}
if the arbitrary function $H$  satisfies the condition
\begin{equation}\label{coord_xform_condition}
\frac{\partial H(r_*,\theta_*)}{\partial r_*}=-\omega_B.
\end{equation}
Substitution of the coordinate transforms (\ref{coord_xform1})
and (\ref{coord_xform2}) into Eq. (\ref{PI-metric}) results in
\begin{equation}
ds^2=  \frac{( -du^2-2du dr_*) \Delta }{ R^2 }  +4 \frac{L^2 sin^2 \theta_* cos^2\theta_* d\theta_*^2 }{ R^2 }   + (\omega_B du + \frac{\partial{H(r_*,\theta_*)}}{\partial{\theta_*}}d\theta_* -d\phi_*)^2 R^2 sin^2\theta.
\end{equation}

We now need to obtain an expression for 
$\frac{\partial H(r_*,\theta_*)}{\partial \theta_*}$.
This is non-trivial as $\omega_B$  is given in terms of $r$ and
$\theta$ rather than in terms of $r_*$ and $\theta_*$.
Since $H$ in the coordinate transform (\ref{coord_xform2}) is arbitrary
up to the condition (\ref{coord_xform_condition}), the remaining
$\frac{\partial H}{\partial\theta_*}$ component is still undefined.
We proceed by defining
\begin{equation}\label{Definition_alpha}
H(r_*,\theta_*)=\int_{r}^{\infty}{\frac{2mas}{\Delta(s) Q(s,\theta_*)}ds} .
\end{equation}
It then follows that Eq. (\ref{coord_xform_condition}) is met, since
\begin{equation}
\frac{\partial{H(r_*,\theta_*)}}{\partial{r_*}}=-\frac{2mar}{\Sigma R^2}=-\omega_B.
\end{equation}
After integration of (\ref{Definition_alpha}),
$\frac{\partial H(r_*,\theta_*)}{\partial \theta_*}$ is obtained as 
\begin{equation}\label{diff_H_theta_*} 
\frac{\partial H(r_*,\theta_*)}{\partial \theta_*}=\omega_B
          \beta \sin(2\theta_*)
\end{equation}
where
\begin{equation}
\beta=\mu   P^2  + \frac{a^3 m}{\omega_B } \int_{r}^{\infty}{\frac{s}{Q^3}ds}.
\label{e-def-beta}
\end{equation}
Note that the integral in Eq.~(\ref{e-def-beta}) is proportional to
$T_1$ defined in Eq.~(\ref{mu_integral term}) and, as remarked earlier,
will need to be evaluated numerically.

The metric now becomes
\begin{equation}
\label{Bondi-Sachs form of the Kerr-metric-2}
ds^2=  \frac{( -du^2-2du dr_*) \Delta }{ R^2 }  +4 \frac{L^2 sin^2 \theta_*
cos^2\theta_* d\theta_*^2 }{ R^2 }   + (\omega_B 
\left[ du + \beta \sin(2\theta_*) d\theta_* \right] -d\phi_*)^2
R^2 sin^2\theta
\end{equation}
which is the Kerr metric in Bondi-Sachs form. 

\subsection{Regularity on the axis $\theta_*=0$}
The situation is very similar to that discussed earlier for
the PI metric in Sec.~\ref{BV_metric4smallTheta}. The only new metric
function that has been introduced is $\beta$, which evaluates on the
axis to
\begin{equation}
\beta=\frac{a^2(5 r^2+a^2)}{8r(r^2+a^2)}.
\end{equation}
The extra terms introduced into the metric components by
way of $\beta$ are multipled by $\theta_*^3$, and are therefore
too small to affect the issue of regularity.

\section{Numerical method and results}
\label{s-n}

\subsection{Coding issues}
We mention only matters that are not routine. 
The expressions containing elliptic integrals $E$ and $F_E$ 
were evaluated using Bulirsch's elliptical numerical solvers \cite{bulirsch-1}. 
The quantity $\Xi(z,k)$ defined in Eq.~(\ref{IP_}) 
can be expressed as a linear combination  of elliptic integrals of various kinds, 
also evaluated using  Bulirsch elliptical numerical solvers \cite{bulirsch-1}. The
quantity $I_2$ is defined in Eq.~(\ref{I2}) by means of an integral whose integrand
decreases monotonically as $\zeta \rightarrow \infty$, therefore standard solvers 
for infinite boundaries can be  used. In similar fashion the first integral in 
Eq.~(\ref{Quasi_Spherical_Surface_eq}) and the integral in Eq.~(\ref{mu_integral term}) 
were evaluated using the same Romberg integration methods.

The quantity $\mu$ is singular as $\theta_* \rightarrow 0$, and this proves
to be problematic for accurate numerical evaluation of metric quantities
near the axis. It is for this reason that, in the Tables below, the smallest
value of $\theta_*$ that is used is 0.3.

\subsection{Construction of data on a regular grid} 
In order to obtain numerically the Ricci tensor of the  metric to $O(h^2)$, 
a 13-point regular grid was used. First we find first metric derivatives
using the 
central difference formula, then we calculate the Christoffel symbols, then 
we calculate first derivatives of the Christoffel symbols using the central 
difference formula, and finally we calculate the various components of the 
Ricci tensor. 
 
For given $(r_{*},\theta_*)$ we obtain the Boyer-Lyndquist coordinates 
$(r,\theta)$ as follows. We make an estimate for $r$ and use 
Eq.~(\ref{Quasi_Spherical_Surface_eq}) to find $\theta$. 
Once $\theta$ has been obtained in this manner, $(r,\theta,\theta_*)$ is 
substituted into Eq.~(\ref{eq-PI-15}) establishing a value for $r_*$. Of course, 
in general this will not be the desired value of $r_*$, so we repeat the 
calculation with a different estimate of $r$, and then use the bisection 
method to find, to an accuracy of $10^{-12}$ the value of $r$ which leads 
to the correct value of $r_*$. 
Now, having obtained $(r,\theta)$ for given $(r_*,\theta_*)$, we find all the 
metric coefficients, including $L$, at the point $(r_*,\theta_*)$. 
 
\subsection{Results}

First, the numerical code was validated by using it to find the Ricci tensor
of the Kerr metric in standard Boyer-Linquist coordinates. Second order convergence to zero was observed. 

The Ricci tensor of the metric (\ref{Bondi-Sachs form of the Kerr-metric-2})
was evaluated.
Tables \ref{tab:r}, \ref{tab:theta}, \ref{tab:L} and \ref{tab:beta_BV1}
show values for
$r$, $\theta$, $L$ and $\beta$ on a regular $(r_*,\theta_*)$
grid centered at $(0.4,0.3)$ with  $m=1$, $a=0.1$, and 
$h=\Delta r_*=\Delta \theta_*=0.01$.
From this data it is straightforward to find numerical values for
all components of the metric
$g_{\alpha\beta}$ on the regular $(r_*,\theta_*)$ grid, and thus to
compute the Ricci tensor $R_{\alpha\beta}$ at the centre of the grid.
The results are shown in the second column of Table~\ref{tab:R_BV1}. Also
in that Table we show $R_{\alpha\beta}$ evaluated using coarser grids,
$h=0.02$ and $h=0.04$, centered at the same point $(0.4,0.3)$
(but in these cases we do not give the intermediate data $r$, $\theta$,
$L$ or $\beta$).
The last columns of the Table show the result of convergence testing,
and it is clear that all components are second order convergent to zero. 

The values of $(r,\theta)$ and of $g_{\alpha\beta}$ were also found
for the cases $m=1$, $a\in\{0.1, 0.2\}$
on regular grids $(r_*,\theta_*)$ centered at $r_*\in\{0.4,0.5\}5$, 
$\theta_*\in\{0.3, 0.6, 0.9, 1.2\}$, and for 
three different discretization parameters namely $h=0.01$, $h=0.02$
and $h=0.04$. Rather than present the convergence rates of all
the components of the 16 sets of Ricci tensors, we instead calculated
the $L_2$-norm defined by
\begin{equation}
\|R_{\alpha\beta}\|=\sqrt{\sum_{\alpha=0}^3\sum_{\beta=0}^3
\frac{(R_{\alpha\beta})^2}{16}}.
\end{equation}
These results are given in Table \ref{tab:L-R_BV1}. In all cases, second-order
convergence to zero was observed.

Because the metric (\ref{Bondi-Sachs form of the Kerr-metric-2}) is
derived from the PI metric, the above results imply that the PI
metric is also Ricci flat; we have performed computations similar
to those described above and explicitly confirmed this -- see
Table \ref{tab:L-R}.

\section{The metric of Fletcher and Lun}
\label{FletcherLun-metric}
Using coordinates $(u,r,\theta,\phi)$ Fletcher and Lun \cite{lun} have
obtained a metric in Bondi-Sachs form, denoted here by FL.
The metric coefficients do not contain implicit functions, so a numerical
evaluation of the Ricci tensor
is straightforward, and convergence to zero was confirmed (see Table
\ref{tab:L-R_Lun}). However, the metric suffers from an important
restriction in that it is not regular at $\theta=0$.

The FL metric coefficients at $\theta=0$ were found to be
\begin{eqnarray}\label{lun_metric_in_leading_order_of_theta}
l_{{tt}}          &=&    {\frac {-  {r}^{2}-  {a}^{2}+{a}^{2} \tanh^2 \alpha  + 2mr}{-  {r}^{2}-  {a}^{2}+{a}^{2} \tanh^2 \alpha }}   \nonumber \\
l_{{tr}}          &=&   -  {\frac {-  {r}^{2}-  {a}^{2}+{a}^{2} \tanh^2 \alpha }{\sqrt {r \left( {r}^{3}+r{a}^{2}+ 2{a}^{2}m \right) }}}     \nonumber \\
l_{{t\theta}}     &=&    {\frac { \left( {r}^{2}  \cosh^2 \alpha- 2mr  \cosh^2 \alpha+{a}^{2} \right) a}{ \left( {r}^{2}  \cosh^2 \alpha+{a}^{2} \right)   \cosh^2 \alpha}}    \nonumber \\
l_{{t\phi}}       &=&   - 2{\frac {amr \tanh^2 \alpha }{-  {r}^{2}-  {a}^{2}+{a}^{2} \tanh^2 \alpha }}    \nonumber \\
l_{{\theta\theta}}&=&   -  {\frac { \left( {r}^{3}  \cosh^2 \alpha+ 2{a}^{2}m+r{a}^{2} \right) r}{ \left( {r}^{2}  \cosh^2 \alpha+{a}^{2} \right)   \cosh^2 \alpha}}    \nonumber \\
l_{{\theta\phi}}  &=&    2{\frac {mr{a}^{2} \sinh^{2}\alpha }{ \left( {r}^{2}  \cosh^2 \alpha+{a}^{2} \right)  \cosh^2 \alpha}}   \nonumber \\ 
l_{{\phi\phi}}    &=&   -  {\frac { \sinh ^2 \alpha \left( {r}^{4}  \cosh^2 \alpha+{r}^{2}{a}^{2}  \cosh^2 \alpha+ 2mr  \cosh^2 \alpha{a}^{2}+{r}^{2}{a}^{2}- 2mr{a}^{2}+{a}^{4} \right)  }{ \left( {r}^{2}  \cosh^2 \alpha+{a}^{2} \right)   \cosh^2 \alpha}}   \nonumber \\
\end{eqnarray}
where $\alpha=\alpha(r,a,m)$ is defined by
\begin{equation}
\alpha=-a\int_r^\infty{\frac{d\nu}{\sqrt{\nu^4+a^2\nu^2+2a^2m\nu}}}.
\end{equation}
In general, all the angular components of the FL metric are non-zero
at $\theta=0$, so the metric cannot be regular there. Now,~\cite{lun}
suggests that the axis of symmetry is not at $\theta=0$ but its location
is a function of $r$, explicitly $\theta=\arcsin (-\tanh(\alpha))$. We do
not comment on the analytic validity of such a statement. Our
aim is to find a form for the Kerr metric that can be used in
numerical codes based on the Bondi-Sachs metric, and whatever the case --
irregularity at $\theta=0$, or an axis of symmetry that is not a straight
coordinate line -- the metric of~\cite{lun} is unsuitable for this purpose.

\section{Conclusion}
\label{s-c}
We have computed numerically the Ricci tensors of the metrics
(\ref{Bondi-Sachs form of the Kerr-metric-2}), PI and FL.
In all cases, second order convergence to zero was observed.
We also investigated regularity near the axis $\theta_*=0$ (or $\theta=0$
in the FL case), finding that the metrics
(\ref{Bondi-Sachs form of the Kerr-metric-2}) 
and PI were regular, but that FL was not.

The aim of this work is to find a metric in Bondi-Sachs form that
represents the Kerr geometry and that can be used in a numerical computation.
The metric (\ref{Bondi-Sachs form of the Kerr-metric-2}) partially
meets that goal. In order to meet it completely, the radial coordinate
$r_*$ will need to be transformed to an area coordinate say $r_{**}$
(i.e., so that the area of the 2-surface $u=r_{**}=\mbox{constant}$ is
$4\pi r^2_{**}$); the procedure for doing so has been described
in~\cite{vishbook}. Also, there will need to be a transformation to
the angular coordinates (stereographic) actually used in the codes.
Finally, the numerical difficulty of finding the metric coefficients
near the axis will need to be overcome. While not trivial, this is
all in principle straightforward, and is deferred to further work.

\section*{Acknowledgments}
We wish to thank Jeffrey Winicour for discussion; and Frans Pretorius for
looking at an earlier draft of the paper, and informing us about an
inconsistency in the calculation of a norm. NTB thanks Max-Planck-Institut
f\"ur GravitationsPhysik, Albert-Einstein-Instit\"ut, for hospitality. 
LRV thanks Cornelius Du Toit, Michael van Canneyt for assistance with the Lazarus and ppc386 compilers.
The work was supported in part by the National Research Foundation,
South Africa, under Grant number 2053724.

\begin{table}[htbp] 
\caption{$r$ on a regular $(r_*=0.4 + ih,\theta_*=0.3 + jh)$ grid
with $h=0.01$, $m=1$ and $a=0.1$; used in the metrics
(\ref{Bondi-Sachs form of the Kerr-metric-2}) and PI.} 
\begin{tabular}{rcccccc} 
\hspace{0.0cm} & $i$    & -2    &  -1    &  0     & 1   & 2 \\ 
$j$ & & & & & \\ 
-2 & &  &  & 2.3701116298499  & &  \\ 
-1 & &  & 2.3685360418458 & 2.3701098379124 & 2.3716892173938 & \\ 
 0 &  & 2.3669659799190 & 2.3685341996528 & 2.3701079904032 & 2.3716873645652 & 2.3732723345139\\ 
1 & &  & 2.3685323027830 & 2.3701060880593 & 2.3716854567442 & \\ 
2 & &  &  & 2.3701041316400 & & 
\end{tabular} 
\label{tab:r} 
\end{table} 

\begin{table}[htbp] 
\caption{$\theta$ on a regular $(r_*=0.4 + ih,\theta_*=0.3 + jh)$ grid
with $h=0.01$, $m=1$ and $a=0.1$; used in the metrics
(\ref{Bondi-Sachs form of the Kerr-metric-2}) and PI.} 
\begin{tabular}{rcccccc} 
\hspace{0.0cm} & $i$    & -2    &  -1    &  0     & 1   & 2 \\ 
$j$  & & & & & \\ 
-2 & &  &  & 0.2797639049518 & &  \\ 
-1 & &  & 0.2897560966969 & 0.2897564200788 & 0.2897567439622  & \\ 
 0 & & 0.2997483666777  & 0.2997486993479 & 0.2997490325375 & 0.2997493662436  & 0.2997497004638\\ 
1 & &  & 0.3097414024170 &  0.3097417452810 & 0.3097420886767 & \\ 
2 & &  &  & 0.3197345612229 & & 
\end{tabular} 
\label{tab:theta} 
\end{table} 
  
\begin{table}[htbp] 
\caption{$L$ on a regular $(r_*=0.4 + ih,\theta_*=0.3 + jh)$ grid
with $h=0.01$, $m=1$ and $a=0.1$; used in the metrics
(\ref{Bondi-Sachs form of the Kerr-metric-2}) and PI.} 
\begin{tabular}{rcccccc} 
\hspace{0.0cm} & $i$       & -2    &  -1    &  0     & 1   & 2 \\ 
$j$ & & & & & & \\ 
-2 &  & &  & 10.5859871980039  & &  \\ 
-1 &  & & 10.2472057969004 & 10.2608136646930 & 10.2744788974703 & \\ 
 0 & & 9.9325339379316 & 9.945685737941 & 9.9588930170935 & 9.9721559726835  
& 9.9854748024450\\ 
1 &  & & 9.6651644706586 & 9.6779990517326 & 9.6908877382673 & \\ 
2 &  & &  & 9.4161849439047 & & 
\end{tabular} 
\label{tab:L}
\end{table} 

\begin{table}[htbp] 
\caption{$\frac{\partial H(r_*.\theta_*)}{\partial \theta_*}=\omega_B\beta$
on a regular $(r_*=0.4 + ih,\theta_*=0.3 + jh)$ grid
with $h=0.01$, $m=1$ and $a=0.1$; used in the metric
(\ref{Bondi-Sachs form of the Kerr-metric-2}).} 
\begin{tabular}{rcccccc} 
\hspace{0.0cm} & $i$    & -2    &  -1    &  0     & 1   & 2 \\ 
$j$  & & & & & \\ 
-2 & &  &  & 0.0000209409007 & &  \\ 
-1 & &  & 0.0000216623673 & 0.0000216050272 & 0.0000215476741 & \\ 
 0 & & 0.0000223786889  & 0.0000223196230 & 0.0000222605430 & 0.0000222014497  & 0.0000221423438\\ 
1 & &  & 0.0000229679828 &  0.0000229071865 & 0.0000228463764 & \\ 
2 & &  &  & 0.0000235446994 & & 
\end{tabular} 
\label{tab:beta_BV1} 
\end{table}

\begin{table} 
\caption{The components of $R_{\alpha\beta}$ 
for the metric (\ref{Bondi-Sachs form of the Kerr-metric-2}) 
at $(r_*=0.4,\theta_*=0.3)$ when $m=1$,
$a=0.1$ for the 3 values of $h$ shown. The data is fitted to
$R_{\alpha\beta}[h]=R_{\alpha\beta}[0]+h^n B_{\alpha\beta}$ and the values of
the 3 unknowns $n$ (the convergence rate, shown below), $R_{\alpha\beta}[0]$
(the convergence limit, shown below), and $B_{\alpha\beta}$ (not shown) can
then be found. In the last column we assume that the convergence limit
is zero, and
the data from the two smallest values of $h$ is fitted to
$R_{\alpha\beta}[h]=h^\rho C_{\alpha\beta}$; then the convergence rate
$\rho$ is found and shown below.} 
\begin{center} 
\begin{tabular}{ccccccc} 
$R_{\alpha\beta}$ & $h=0.01$ & $h=0.02$ & $h=0.04$ & $n$ & $R_{\alpha\beta}[0]$
& $\rho$ \\
$R_{00}$ &  -.0000000380807   &   -.0000001511411  & -.0000006048914  & 2.0048  & -.0000000005609 & 1.9888\\ 
$R_{01}$ &  -.0000000356994   &   -.0000001416176  & -.0000005668227  & 2.0052  & -.0000000005629 & 1.9880\\ 
$R_{02}$ &   .0000000002564   &    .0000000010245  &  .0000000040897  & 1.9966  & -.0000000000005 & 1.9985\\ 
$R_{03}$ &   .0000001685141   &    .0000006739409  &  .0000026940829  & 1.9989  & -.0000000001356 & 1.9998\\ 
$R_{10}$ &  -.0000000356994   &   -.0000001416176  & -.0000005668227  & 2.0052  & -.0000000005629 & 1.9880\\ 
$R_{11}$ &   .0000001077265   &    .0000004299945  &  .0000017202702  & 2.0013  &  .0000000004374 & 1.9969\\ 
$R_{12}$ &  -.0000000012202   &   -.0000000044084  & -.0000000175449  & 2.0428  & -.0000000001985 & 1.8531\\ 
$R_{13}$ &   .0000000005647   &    .0000000022268  &  .0000000089078  & 2.0071  &  .0000000000143 & 1.9794\\ 
$R_{20}$ &   .0000000002564   &    .0000000010245  &  .0000000040897  & 1.9966  & -.0000000000005 & 1.9985\\ 
$R_{21}$ &   .0000003176420   &    .0000012706415  &  .0000050765232  & 1.9977  & -.0000000007055 & 2.0001\\ 
$R_{22}$ &   .0129878554048   &    .0521063939839  &  .2109476828529  & 2.0217  &  .0002061572158 & 2.0043\\
$R_{23}$ &  -.0000000137600   &   -.0000000550017  & -.0000002195263  & 1.9961  &  .0000000000366 & 1.9990\\ 
$R_{30}$ &   .0000001685141   &    .0000006739409  &  .0000026940829  & 1.9989  & -.0000000001356 & 1.9998\\ 
$R_{31}$ &   .0000000005647   &    .0000000022268  &  .0000000089078  & 2.0071  &  .0000000000143 & 1.9794\\
$R_{32}$ &  -.0000000137600   &   -.0000000550017  & -.0000002195263  & 1.9961  &  .0000000000366 & 1.9990\\
$R_{33}$ &  -.0000111845769   &   -.0000447327548  & -.0001788197801  & 1.9989  &  .0000000099046 & 1.9998 
\end{tabular} 
\label{tab:R_BV1} 
\label{FROM BV-1 data archives:
Data from Ruv_And_L-2Norm_h=0.01_r_dash=0.4_Theta=0.3_m=1_p=1E-13_a=0.1.dat inarchive a_range=0.1 0.2_r_dash_range=0.4 0.5_Theta_range=0.3 0.6 0.9 1.2_h_range=0.01 0.02 0.04.tar.gz Note: L-norms are .00324696505683892535672529648541 .0130266033029653093725796504500 .0527369396876306157638785120116 respectively. This should be first three value entries in table to follow if the data is correctly related. The columns n and resultant Richardson Limit can be found in Maple File Richardson_Limits_4paper.mws }
\end{center} 
\end{table}

\begin{table}[htbp]
\caption{The $L_2$-norm $\|R_{\alpha\beta}\|$ of the metric
(\ref{Bondi-Sachs form of the Kerr-metric-2}) at the gridpoints $(r_*=0.4,0.5,\mbox{  }\theta_*=0.3,0.6,0.9,1.2)$  when $m=1$,
$a=0.1,0.2$ for the 3 values of $h$ shown. The data is fitted to
$\|R_{\alpha\beta}[h]\|=\|R_{\alpha\beta}[0]\|+h^n B_{\alpha\beta}$
and the values of the 3 unknowns $n$ (the convergence rate, shown below),
$\|R_{\alpha\beta}[0]\|$ (the convergence limit, shown below), and
$B_{\alpha\beta}$ (not shown) can then be found. In the last column we assume that the convergence limit is zero, and
the data from the two smallest values of $h$ is fitted to
$R_{\alpha\beta}[h]=h^\rho C_{\alpha\beta}$; then the convergence rate
$\rho$ is found and shown below} 
\begin{center} 
\begin{tabular}{ccccccccccc}
$a$&$r*$&$\theta_*$ & $ $ & $h=0.01$ & $h=0.02$ & $h=0.04$ & $n$ & $\|R_{\alpha\beta}[0]\|$ & $\rho$   & \\

$0.1$&$0.4$&$0.3$ &  $|$ & .0032469650568 & .0130266033030 & .0527369396876 & 2.0217 &  .0000515392842 & 2.0043 &   \\ 
$0.1$&$0.4$&$0.6$ &  $|$ & .0002130833763 & .0008527906803 & .0034184910131 & 2.0038 &  .0000006083421 & 2.0008 &   \\ 
$0.1$&$0.4$&$0.9$ &  $|$ & .0000389028104 & .0001556196190 & .0006226264251 & 2.0004 &  .0000000127093 & 2.0001 &   \\ 
$0.1$&$0.4$&$1.2$ &  $|$ & .0000293455458 & .0001173591202 & .0004691359914 & 1.9988 & -.0000000231698 & 1.9997 &   \\ 
$0.1$&$0.5$&$0.3$ &  $|$ & .0032470081286 & .0130267772029 & .0527376265713 & 2.0217 &  .0000515376608 & 2.0043 &   \\ 
$0.1$&$0.5$&$0.6$ &  $|$ & .0002131086500 & .0008528867283 & .0034188769168 & 2.0039 &  .0000006107702 & 2.0008 &   \\ 
$0.1$&$0.5$&$0.9$ &  $|$ & .0000389167189 & .0001556755506 & .0006228548385 & 2.0004 &  .0000000130975 & 2.0001 &   \\ 
$0.1$&$0.5$&$1.2$ &  $|$ & .0000293127899 & .0001172283812 & .0004686121003 & 1.9989 & -.0000000234006 & 1.9997 &   \\  
$0.2$&$0.4$&$0.3$ &  $|$ & .0032453323914 & .0130201489364 & .0527123284154 & 2.0217 &  .0000516338458 & 2.0043 &   \\ 
$0.2$&$0.4$&$0.6$ &  $|$ & .0002143149911 & .0008577236053 & .0034382789181 & 2.0038 &  .0000006116897 & 2.0008 &   \\ 
$0.2$&$0.4$&$0.9$ &  $|$ & .0000393784101 & .0001575247969 & .0006302707322 & 2.0005 &  .0000000140940 & 2.0001 &   \\ 
$0.2$&$0.4$&$1.2$ &  $|$ & .0000298331343 & .0001193103237 & .0004769318246 & 1.9988 & -.0000000245470 & 1.9997 &   \\ 
$0.2$&$0.5$&$0.3$ &  $|$ & .0032454107976 & .0130204532491 & .0527135097953 & 2.0217 &  .0000516341443 & 2.0043 &   \\ 
$0.2$&$0.5$&$0.6$ &  $|$ & .0002143150084 & .0008577193834 & .0034382574117 & 2.0039 &  .0000006130839 & 2.0008 &   \\ 
$0.2$&$0.5$&$0.9$ &  $|$ & .0000393833905 & .0001575427252 & .0006303423768 & 2.0004 &  .0000000149722 & 2.0001 &   \\ 
$0.2$&$0.5$&$1.2$ &  $|$ & .0000297893272 & .0001191387256 & .0004762466973 & 1.9988 & -.0000000260206 & 1.9998 &   \\
\end{tabular} 
\label{tab:L-R_BV1} 
\end{center} 
\end{table} 

\begin{table}[htbp] 
\caption{The $L_2$-norm $\|R_{\alpha\beta}\|$ for the PI metric at the
grid points $(r_*=0.4,0.5,\mbox{ }\theta_*=0.3,0.6,0.9,1.2)$  when $m=1$,
$a=0.1,0.2$ for the 3 values of $h$ shown. The data is fitted to
$\|R_{\alpha\beta}[h]\|=\|R_{\alpha\beta}[0]\|+h^n B_{\alpha\beta}$ and the values of the 3 unknowns $n$ (the convergence rate, shown below),
$\|R_{\alpha\beta}[0]\|$ (the convergence limit, shown below), and
$B_{\alpha\beta}$ (not shown) can then be found. In the last column we assume that the convergence limit is zero, and
the data from the two smallest values of $h$ is fitted to
$R_{\alpha\beta}[h]=h^\rho C_{\alpha\beta}$; then the convergence rate
$\rho$ is found and shown below}
\begin{center} 
\begin{tabular}{ccccccccccc}
$a$&$r*$&$\theta_*$ & $ $ & $h=0.01$ & $h=0.02$ & $h=0.04$ & $n$ & $\|R_{\alpha\beta}[0]\|$ & $\rho$ \\

$0.1$&$0.4$&$0.3$ &  $|$ & .0032469652386 &  .0130266040297 & .0527369425946  &  2.0217&   .0000515392845 & 2.0043 &   \\ 
$0.1$&$0.4$&$0.6$ &  $|$ & .0002130840309 &  .0008527932981 & .0034185014848  &  2.0039&   .0000006083423 & 2.0008 &   \\ 
$0.1$&$0.4$&$0.9$ &  $|$ & .0000389034943 &  .0001556223556 & .0006226373828  &  2.0004&   .0000000127102 & 2.0001 &   \\  
$0.1$&$0.4$&$1.2$ &  $|$ & .0000293439533 &  .0001173527506 & .0004691105137  &  1.9988&  -.0000000231700 & 1.9997 &   \\  
$0.1$&$0.5$&$0.3$ &  $|$ & .0032470083035 &  .0130267779027 & .0527376293706  &  2.0217&   .0000515376608 & 2.0043 &   \\  
$0.1$&$0.5$&$0.6$ &  $|$ & .0002131092800 &  .0008528892486 & .0034188869986  &  2.0039&   .0000006107701 & 2.0008 &   \\  
$0.1$&$0.5$&$0.9$ &  $|$ & .0000389173773 &  .0001556781866 & .0006228653942  &  2.0004&   .0000000130977 & 2.0001 &   \\  
$0.1$&$0.5$&$1.2$ &  $|$ & .0000293112574 &  .0001172222518 & .0004685875840  &  1.9989&  -.0000000234007 & 1.9997 &   \\   
$0.2$&$0.4$&$0.3$ &  $|$ & .0032453330980 &  .0130201517634 & .0527123397244  &  2.0217&   .0000516338460 & 2.0043 &   \\  
$0.2$&$0.4$&$0.6$ &  $|$ & .0002143175794 &  .0008577339580 & .0034383203299  &  2.0039&   .0000006116900 & 2.0008 &   \\  
$0.2$&$0.4$&$0.9$ &  $|$ & .0000393812217 &  .0001575360447 & .0006303157634  &  2.0005&   .0000000140979 & 2.0001 &   \\  
$0.2$&$0.4$&$1.2$ &  $|$ & .0000298268109 &  .0001192850311 & .0004768306609  &  1.9988&  -.0000000245469 & 1.9997 &   \\  
$0.2$&$0.5$&$0.3$ &  $|$ & .0032454114777 &  .0130204559705 & .0527135206818  &  2.0217&   .0000516341442 & 2.0043 &   \\  
$0.2$&$0.5$&$0.6$ &  $|$ & .0002143174996 &  .0008577293475 & .0034382972686  &  2.0039&   .0000006130844 & 2.0008 &   \\  
$0.2$&$0.5$&$0.9$ &  $|$ & .0000393860967 &  .0001575535526 & .0006303857252  &  2.0005&   .0000000149752 & 2.0001 &   \\  
$0.2$&$0.5$&$1.2$ &  $|$ & .0000297832468 &  .0001191144036 & .0004761494153  &  1.9988&  -.0000000260198 & 1.9998 &   \\
\end{tabular} 
\label{tab:L-R} 
\end{center} 
\end{table} 

\begin{table}[htbp]
\caption{The $L_2$-norm $\|R_{\alpha\beta}\|$ for the FL metric at the grid points
$(r=4,5 ,\mbox{  }\theta=0.3,0.6,0.9,1.2)$  when $m=1$,
$a=0.1,0.2$ for the 3 values of $h$ shown. The data is fitted to
$\|R_{\alpha\beta}[h]\|=\|R_{\alpha\beta}[0]\|+h^n B_{\alpha\beta}$ and the values of
the 3 unknowns $n$ (the convergence rate, shown below), $\|R_{\alpha\beta}[0]\|$
(the convergence limit, shown below), and $B_{\alpha\beta}$ (not shown) can
then be found. In the last column we assume that the convergence limit is zero, and
the data from the two smallest values of $h$ is fitted to
$R_{\alpha\beta}[h]=h^\rho C_{\alpha\beta}$; then the convergence rate
$\rho$ is found and shown below} 
\begin{center} 
\begin{tabular}{ccccccccccc}
$a$&$r$&$\theta$ & $ $ & $h=0.001$ & $h=0.002$ & $h=0.004$ & $n$ & $\|R_{\alpha\beta}[0]\|$ & $\rho$  \\

$0.1$&$4$&$0.3$ &  $|$ & .0000457585190 & .0001830406822 & .0007322693156    &  2.0003&  .0000000089023 & 2.0001 & \\ 
$0.1$&$4$&$0.6$ &  $|$ & .0000025606471 & .0000102426033 & .0000409714183    &  2.0000&  .0000000001052 & 2.0000 & \\ 
$0.1$&$4$&$0.9$ &  $|$ & .0000004382731 & .0000017531700 & .0000070127199    &  2.0000& -.0000000000300 & 2.0001 & \\ 
$0.1$&$4$&$1.2$ &  $|$ & .0000002993400 & .0000011973612 & .0000047894342    &  2.0000& -.0000000000017 & 2.0000 & \\ 
$0.1$&$5$&$0.3$ &  $|$ & .0000426162736 & .0001704709495 & .0006819789861    &  2.0003&  .0000000079718 & 2.0000 & \\ 
$0.1$&$5$&$0.6$ &  $|$ & .0000024663873 & .0000098656545 & .0000394635674    &  2.0000&  .0000000000586 & 2.0000 & \\ 
$0.1$&$5$&$0.9$ &  $|$ & .0000004274473 & .0000017097918 & .0000068391881    &  2.0000&  .0000000000012 & 2.0000 & \\ 
$0.1$&$5$&$1.2$ &  $|$ & .0000002969713 & .0000011878357 & .0000047512974    &  2.0000&  .0000000000169 & 1.9999 & \\  
$0.2$&$4$&$0.3$ &  $|$ & .0000666096949 & .0002664508787 & .0010659963373    &  2.0003&  .0000000160412 & 2.0001 & \\ 
$0.2$&$4$&$0.6$ &  $|$ & .0000030950589 & .0000123802850 & .0000495225364    &  2.0000&  .0000000001332 & 2.0000 & \\ 
$0.2$&$4$&$0.9$ &  $|$ & .0000004998811 & .0000019995370 & .0000079982078    &  2.0000&  .0000000000011 & 2.0000 & \\ 
$0.2$&$4$&$1.2$ &  $|$ & .0000003097359 & .0000012388778 & .0000049554632    &  2.0000&  .0000000000239 & 1.9999 & \\ 
$0.2$&$5$&$0.3$ &  $|$ & .0000570798645 & .0002283289638 & .0009134670723    &  2.0003&  .0000000125727 & 2.0001 & \\ 
$0.2$&$5$&$0.6$ &  $|$ & .0000028663076 & .0000114652931 & .0000458624019    &  2.0000&  .0000000001087 & 2.0000 & \\ 
$0.2$&$5$&$0.9$ &  $|$ & .0000004733989 & .0000018936611 & .0000075746822    &  2.0000& -.0000000000248 & 2.0000 & \\ 
$0.2$&$5$&$1.2$ &  $|$ & .0000003046906 & .0000012187660 & .0000048750369    &  2.0000& -.0000000000045 & 2.0000 & \\ 
\end{tabular} 
\label{tab:L-R_Lun} 
\end{center} 
\end{table}


\begin{thebibliography}{99}

\bibitem{lehner} L. Lehner, Class. Quantum Grav. {\bf 18}, R25 (2001).

\bibitem{cook} G. Cook, Living Rev. Relativity,
http://www.livingreviews.org/lrr-2000-5 (2000).

\bibitem{bondi}
H. Bondi, M.J.G. van der Burg and A.W.K. Metzner,
Proc. R. Soc. {\bf A269}, 21 (1962).

\bibitem{sachs}
R.K. Sachs, Proc. R. Soc. {\bf A270}, 103 (1962).

\bibitem{newnews}
N.T. Bishop and S.S. Deshingkar,
Phys. Rev. D {\bf  68}, 024031 (2003).

\bibitem{particle}
N.T. Bishop, R. Gomez, S. Husa, L. Lehner and J. Winicour,
Phys. Rev. D {\bf  68}, 084015 (2003).

\bibitem{roberto}
R. G\'{o}mez, Phys. Rev. D {\bf 64}, 024007 (2001).

\bibitem{rdi2}
R.A. d'Inverno, M.R. Dubal and E.A. Sarkies, Class. Quantum Grav.
{\bf 17}, 3157 (2000).

\bibitem{siebel1} 
F.~Siebel, J.~A.~Font, E.~Muller and P.~Papadopoulos, 
Phys. Rev. D {\bf 67}, 124018 (2003).

\bibitem{bartnik}
R. Bartnik, Class. Quantum Grav. {\bf 14}, 2185 (1997).

\bibitem{winLR}
J. Winicour, Living Rev. Relativity,
http://www.livingreviews.org/lrr-2001-3 (2001).

\bibitem{lun} S.J. Fletcher and A.W.C. Lun, Class. Quantum Grav. {\bf 20}, 4153 (2003).

\bibitem{win} J. Winicour, private communication (2005).

\bibitem{PI-1} F. Pretorius and W. Israel, Class. Quantum Grav {\bf 15}, 2289 (1998). 

\bibitem{bulirsch-1} R. Bulirsch, Numerische Mathematik {\bf 7}, 78 (1965); {\bf 13},
305 (1969).  

\bibitem{vishbook}
N.T. Bishop, R. Gomez, L. Lehner, B. Szilagyi, J. Winicour and
R.A. Isaacson, in B.R. Iyer and B. Bhawal (Eds.) 
{\it Black Holes, Gravitational Radiation and the Universe} (Kluwer Academic
Publishers, Dordrecht, 1999).

\end{thebibliography}
\end{document}